\documentclass[12pt,preprint]{aastex}

\usepackage{graphicx}
\usepackage{tipa}

\shorttitle{Active Asteroid 288P}
\shortauthors{Agarwal, et al.}

\begin{document}

\title{Hubble and Keck Telescope Observations of Active Asteroid 288P/300163 (2006 VW139)}

\author{Jessica Agarwal$^1$, David Jewitt$^{2,3}$, Harold Weaver$^4$, Max Mutchler$^5$, Stephen Larson$^6$ }
\affil{$1$  Max Planck Institute for Solar System Research, Justus-von-Liebig-Weg 3, 37077 G\"ottingen, Germany \\
$2$ Dept. Earth, Planetary and Space Sciences, UCLA, 595 Charles Young Drive East, Box 951567
Los Angeles, CA 90095-1567 \\
$3$ Dept. Physics and Astronomy, UCLA, 430 Portola Plaza, 
        Los Angeles, CA 90095-1547 \\
$4$ The Johns Hopkins University Applied Physics Laboratory, 11100 Johns Hopkins Road, Laurel, Maryland 20723  \\
$5$ Space Telescope Science Institute, 3700 San Martin Drive, Baltimore, MD 21218 \\
$6$ Lunar and Planetary Laboratory, University of Arizona, 1629 E. University Blvd.
Tucson AZ 85721-0092 \\
}
\email{contact: agarwal@mps.mpg.de}

\begin{abstract} 

We present \emph{Hubble Space Telescope} and Keck 10 meter telescope observations of active asteroid 288P/300163 (2006 VW139) taken to examine ejected dust.  The nucleus is a C-type object with absolute magnitude \mbox{$H_V$ = 17.0$\pm$0.1} and estimated diameter $\sim$2.6 km (for assumed visual geometric albedo $p_V$ = 0.04).  
Variations in the brightness of the nucleus at the 10\% to 15\% level are significant in both 2011 December and 2012 October but we possess too few data to distinguish variations caused by activity from those caused by rotation.  The dust scattering cross-section in 2011 December is  $\sim$40 km$^2$, corresponding to a dust mass $\sim$9$\times$10$^6$ kg (88 $\mu$m mean particle radius assumed).  The full width at half maximum of the debris sheet varies from $\sim$100 km near the nucleus to $\sim$1000 km 30\arcsec (40,000 km) east of it.   Dust dynamical models indicate ejection speeds between 0.06 and 0.3 m s$^{-1}$, particle sizes between 10 and 300 $\mu$m and an inverse square-root relation between particle size and velocity. Overall, the data are most simply explained by prolonged, low velocity ejection of dust, starting in or before  2011 July and continuing until at least 2011 October.  These properties are consistent with the sublimation of near-surface ice aided by centrifugal forces. The high spatial resolution of our HST images (52 km per pixel) reveals details that remained hidden in previous ground-based observations, such as the extraordinarily small vertical extent of the dust sheet, ejection speeds well below the nucleus escape speed, and the possibility of a binary nucleus.

\end{abstract}


\clearpage

\section{INTRODUCTION}

The comet-like appearance of numbered asteroid 300163 (formerly 2006 VW139) was first noticed on  UT 2011 \mbox{August 30} (Hsieh et al.~2012), about six weeks after perihelion (UT 2011 Jul 18.54, at 2.438 AU).  The orbit of this object, since renamed as comet 288P/300163 (hereafter 288P), lies in the outer asteroid belt (semimajor axis, $a$ = 3.050 AU, eccentricity, $e$ = 0.200, and inclination, $i$ = 3.2$\degr$). Its Tisserand parameter, $T_J$ = 3.202, is typical of asteroids and lies far above the $T_J$ = 3 dynamical dividing line separating comets from asteroids (Kresak 1982).    Numerical experiments show that capturing a body with such a large $T_J$ from the Kuiper belt is extraordinarily difficult, involving either the action of sustained non-gravitational forces orders of magnitude larger than observed in comets (Fern{\'a}ndez et al.~2002, Levison et al.~2006) or a dramatic re-arrangement of the planetary orbits (Levison et al.~2009).  Consequently, 288P is classified as one of about twenty presently-known active asteroids (Jewitt et al.~2015c).

Active asteroids are solar system objects characterized by having 1)  orbits interior to Jupiter's, 2)  Tisserand parameters with respect to Jupiter, $T_J >$ 3, and 3)  dust ejected in quantities sufficient to produce a comet-like coma or tail (Jewitt 2012).  These are sometimes labeled main-belt comets (MBCs; Hsieh and Jewitt 2006).   Many different physical mechanisms operate to account for mass loss from the active asteroids (Jewitt 2012).  For example, 100 km -sized (596) Scheila ejected dust following the impact of a $\sim$30 meter sized projectile (Bodewits et al.~2011, Jewitt et al.~2011, Ishiguro et al.~2011, Bodewits et al.~2014). The tiny ($\sim$100 meter diameter) active asteroid P/2010 A2 is either impact-produced or the result of rotational breakup, perhaps driven by radiation torques (Jewitt et al.~2010, Agarwal et al.~2013). Rapid rotation has also been implicated in mass shedding from asteroids 311P/PANSTARRS (formerly P/2013 P5) (Jewitt et al.~2013b, 2015a) and 62412 (Sheppard and Trujillo 2015), in the break-up of asteroid P/2013 R3 into ten or more discrete sub-nuclei (Jewitt et al.~2014a), and possibly in the ejection of fragments from P/2012 F5 (Drahus et al., 2015). Four active asteroids, 133P/Elst-Pizarro (Hsieh et al.~2004), 238P/Read (Hsieh et al.~2011), 313P/Gibbs (Jewitt et al.~2015b,d), and 324P/La Sagra (Hsieh and Sheppard, 2015) show evidence for recurrent activity, which is an expected signature of water ice sublimation. 

Here, we analyse high resolution Hubble Space Telescope (\emph{HST}) images obtained on UT 2011 December 07 and 15. The second observation was timed to occur as the Earth passed through the orbital plane of 288P, providing a unique perspective on the ejected dust. We use these observations in order to determine the vertical extent of the dust released from the asteroid free from the effects of projection. Separately, we obtained supporting observations in search of long-lived dust at the Keck 10-meter telescope on UT 2012 October 14. A journal of observations is provided in Table (\ref{geometry}).

\section{OBSERVATIONS}

\begin{figure}
\epsscale{1.0}
\begin{center}
\plotone{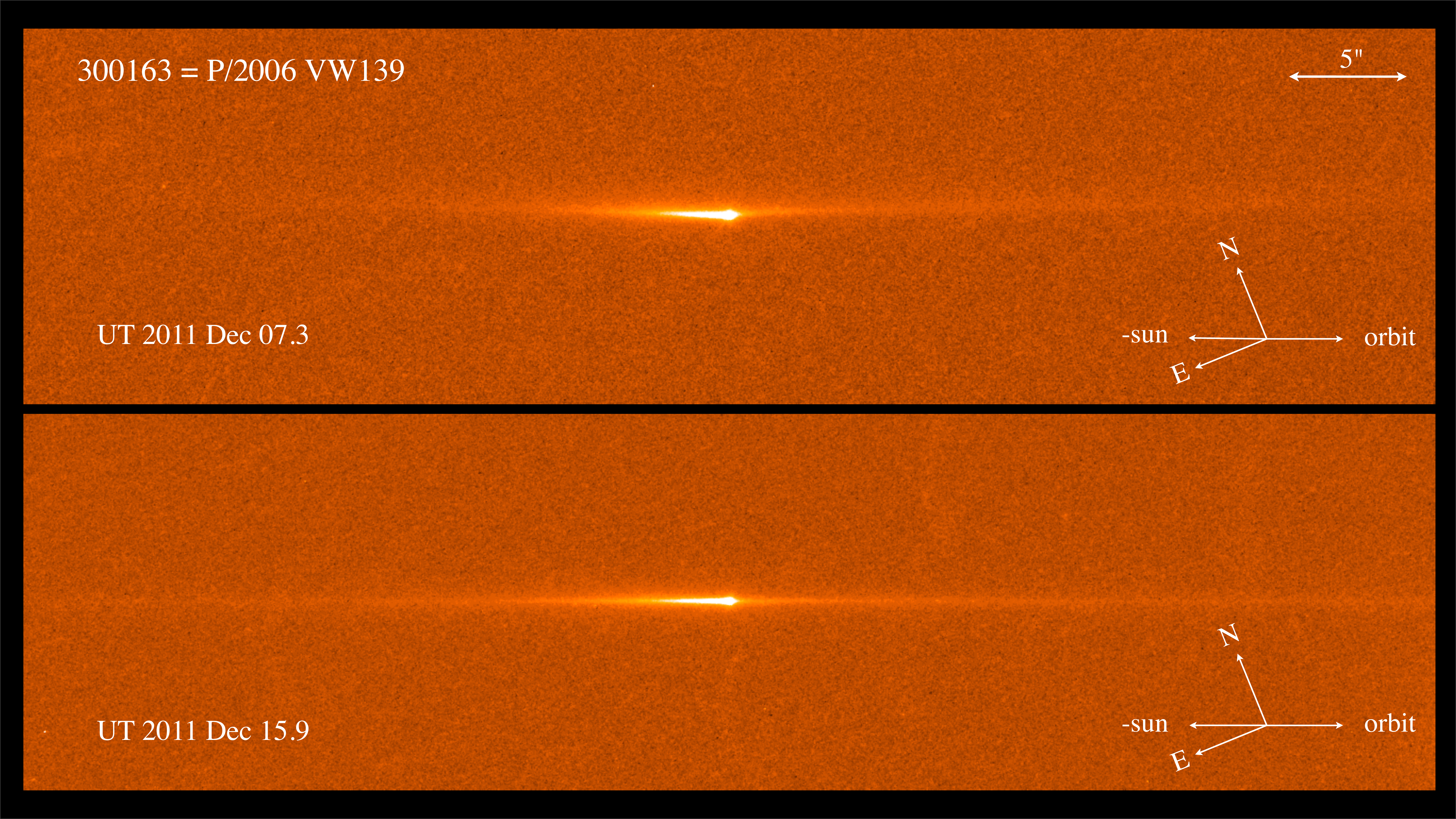}
\caption{Drizzle-combined average images of 288P taken UT December 07 (upper) and December 15 (lower).  Cosmic rays have been removed from the images, each of total integration time 1650 s.  The cardinal directions are marked together with the extended anti-solar vector (labeled ``-sun'')  and the orbit of the asteroid, both projected into the plane of the sky.  Each panel shows a region 80\arcsec~$\times$ 16\arcsec. \label{image_hst} } 
\end{center} 
\end{figure}

We triggered a pre-existing Target of Opportunity program on the Hubble Space Telescope (GO 12597), obtaining data on UT 2011 December~07.253 - 07.289 and UT 2011 December~15.843 - 15.872 with the WFC3 camera~(Dressel 2015). On each date we took four exposures of 350~s duration and one of 250~s to examine the low surface brightness coma.  The  integrations employed the  F606W filter ($\lambda_c \sim$ 6000\AA~and FWHM  $\sim$ 2300~\AA). At the time of observation, the 0.04$\arcsec$ pixels corresponded to 52~km at the comet, so that the Nyquist sampled (2~pixel) resolution was 104 km. Drizzle-combined average images from each date having 1650~s total integration time are shown in Figure (\ref{image_hst}).

Using the Keck 10 meter telescope on Mauna Kea, Hawaii, we obtained a sequence of images in the B, V and R filters on UT 2012 October 14.  The LRIS camera (Oke et al.~1995) offers independent blue and red channels, permitting simultaneous observations in two wavelengths.  We used a dichroic beam-splitter with 50\% transmission near 4900\AA\ to separate the channels.  Integration times were 300 s for the V and R filter data and 340 s for the B filter.  The data were flat fielded using images of an illuminated patch on the inside of the Keck dome.  Photometric calibration was obtained using the stars PG 0918 + 029A and 94-401 from the catalog by Landolt (1992).   All images were obtained with the telescope tracked at non-sidereal rates to follow the motion of 288P.  Seeing varied from about 0.7\arcsec~to 1.1\arcsec~FWHM (corresponding to 1670 km to 2630 km) during the observations.

\section{PHOTOMETRY AND MORPHOLOGY}
\subsection{Nucleus}
\label{subsec:nucleus}
\begin{figure}
\epsscale{0.95}
\begin{center}
\plotone{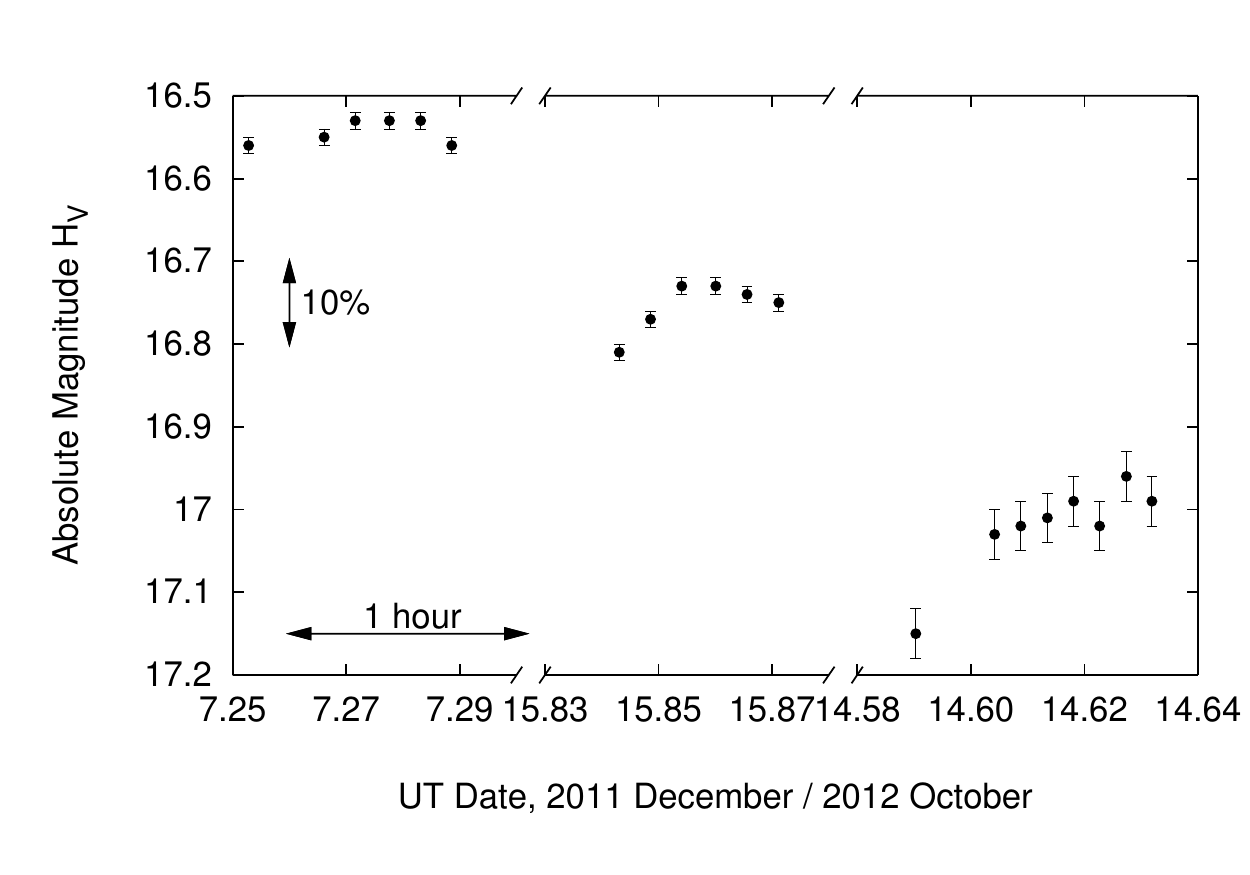}
\caption{Absolute magnitudes derived from photometry of the nucleus of 288P on UT 2011 December 07 and 15, and 2012 October 14 (c.f.~Tables~\ref{photometry_hst} and \ref{photometry_keck}). The errorbars represent only the uncertainty of the individual measurements. In addition, the derived data are subject to a systematic uncertainty of 0.1 magnitudes due to the unknown phase function.\label{fig:absmag} } 
\end{center} 
\end{figure}

In the two HST images, the nucleus is not easily separable from the dust in which it is embedded, even at the Nyquist-sampled $\sim$100 km resolution of Hubble. Table (\ref{photometry_hst}) shows photometry of the near-nucleus region obtained within a circular aperture of projected radius 5 pixels (0.2\arcsec), with background (coma and sky) subtraction from a concentric annulus having inner and outer radii 9 pixels and 16 pixels (0.36\arcsec~and 0.64\arcsec, respectively).  The Table (see also Figure (\ref{fig:absmag})) shows that the nucleus is photometrically variable at the level of $\sim$10\% within each 1-hour HST orbit, whereas the statistical photometric errors are much smaller, $\sim \pm$1\%.  The apparent magnitude also faded by $\sim$0.4 magnitudes in the 8 days between December 07 and 15.  

We computed absolute magnitudes (i.e.~corrected to unit heliocentric and geocentric distances ($R$ = $\Delta$ = 1 AU) and to opposition) from   

\begin{equation}
H_V = m_V - 5\log_{10}(R\Delta) + 2.5\log_{10}(\Phi(\alpha))
\label{absmag}
\end{equation}

\noindent and show the results in  Table (\ref{photometry_hst}) and Figure~(\ref{fig:absmag}). In Equation (\ref{absmag}), $\Phi(\alpha)$ is the phase function correction equal to the ratio of the scattered flux density at phase angle $\alpha$ to that at $\alpha$=0\degr.  Unfortunately, $\Phi(\alpha)$ is unmeasured for 288P.  We use the HG approximation with scattering parameter $g$ = 0.15 as appropriate for C-type asteroids (Bowell et al.~1989) to calculate the absolute magnitudes.  The faintest absolute magnitude from the HST data is $H_V$ = 16.81 on UT 2011 December 15.  If we had instead assumed $g$ = 0.25, more representative of the phase functions of S-type asteroids, the resulting $H_V$ would be fainter by 0.12 magnitudes in the phase angle range in which 288P was observed (Table (\ref{geometry})).  The  $\sim$0.1 magnitude difference between these phase corrections provides our best estimate of the systematic uncertainty in $H_V$ resulting from the un-measured phase function.  

While the apparent brightness faded by $\sim$0.4 magnitudes between the two observations, the mean absolute magnitude faded by only  $\sim$0.2 magnitudes~from $H_V$ = 16.54 on UT 2011 December 7 to $H_V$ = 16.76 on UT 2011 December 15 (Table (\ref{photometry_hst})).  Although modest, this fading is too large to be explained by the uncertainty of the phase function (the phase angle difference is only $\sim$2\degr) and instead suggests the loss of dust from the region within $\sim$250 km of the nucleus. Alternatively, the two observations could have sampled different phases of the rotational lightcurve of an elongated nucleus, if the rotation period is long compared to the 1~hour duration of each HST orbit. Approximating the asteroid as a prolate spheroid, a minimum axis ratio of 1.2 is required to explain a magnitude difference of $\sim$0.2 magnitudes, which is compatible with the shapes of known asteroids.
Absolute magnitudes were also presented in Table 1 of Hsieh et al.~(2012). Their results for UT 2011 December 04 - 16, corrected to the V-filter, are $>$1 magnitude brighter ($H_V$ = 15.6) than in our HST data, obtained nearly simultaneously. As they recognized, this presumably reflects contamination of their large-aperture photometry by near-nucleus dust.

Keck imaging data from UT 2012 October 14 provide additional constraints.  The most striking feature of the Keck data is the apparent absence of dust, even though these are deep observations with the world's largest telescope. The integrated magnitudes and colors of 288P were determined from Keck data using apertures 3.4\arcsec~(25 pixels) in radius with sky subtraction from a contiguous annulus extending to 6.8\arcsec~radius.  We find $m_R$ = 22.45$\pm$0.02, $m_B - m_V$ = 0.67$\pm$0.04 and $m_V - m_R$ = 0.50$\pm$0.03, where the quoted uncertainties are estimates based on the scatter of the data and measurements of other targets of similar brightness.  Hsieh et al.~(2012) reported $m_B - m_R$ = 1.06$\pm$0.04, compared with $m_B - m_R$ = 1.17$\pm$0.04 here.  The difference is probably insignificant, given that neither work fully sampled the rotational lightcurve and that the earlier measurements by Hsieh et al.~on UT 2011 December 04 are more likely to be contaminated by dust than those a year later from Keck.   For comparison, the solar color is $m_B - m_R$ = 1.00$\pm$0.02 (Holmberg et al.~2006). Overall, 288P is slightly redder than the Sun, consistent with having a C-type asteroidal surface, typical of asteroids in its vicinity (Ivezic et al.~2002).  
 
We again computed the absolute magnitude using Equation (\ref{absmag}) and the HG phase function with parameter $g$ = 0.15.  The average value is $H_V$ = 17.0$\pm$0.1 (Table (\ref{photometry_keck}) and Figure~(\ref{fig:absmag})), where the dominant uncertainty results from the phase angle correction, again with variations of order 15\% on the $\sim$1 hr timescale of the measurements probably caused by nucleus rotation. Figure~(\ref{fig:absmag}) shows that the average absolute magnitude in the Keck data faded by an additional 0.24 magnitudes~relative to the second HST data set from a year earlier, again consistent with the progressive loss of dust from the near-nucleus region.

\begin{figure}
\epsscale{1.0}
\begin{center}
\plotone{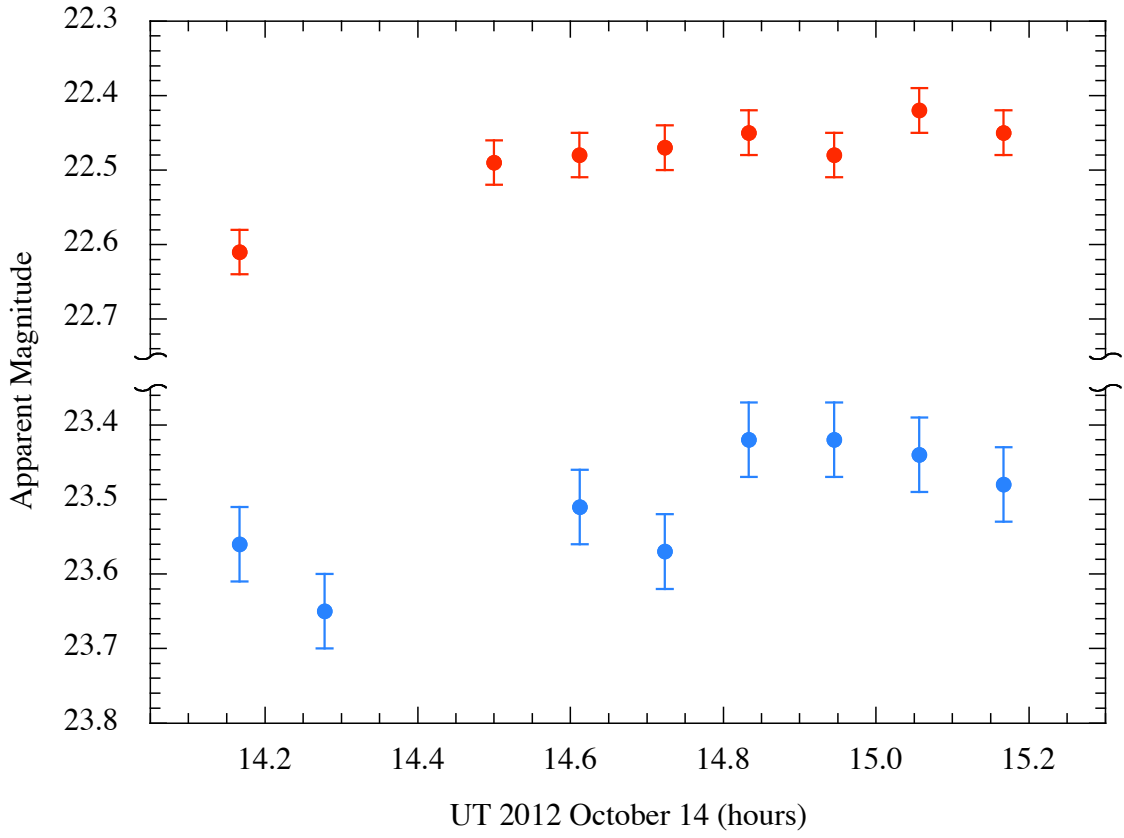}
\caption{Lightcurve on UT 2012 October 14 obtained at the Keck telescope.  Red and blue points refer to measurements through the R and B filters, respectively.   \label{keck_lightcurve} } 
\end{center} 
\end{figure}

We analyzed the individual Keck images in order to search for brightness variations in the object.  We used circular photometry apertures of projected radius 2.03\arcsec~(15 pixels) and made a small correction (typically 0.04 mag) for light lost by the use of this aperture.  The resulting lightcurves are only $\sim$1 hr in duration, but show variations that are large compared to the uncertainties on individual determinations (Table \ref{photometry_keck} and Figure \ref{keck_lightcurve}).  Both the R-band and B-band brightnesses increased by about 0.15 magnitudes~over the interval of observation.  The data are too limited to determine the nucleus rotation period but very short periods ($<$2 hr, double-peaked) of the sort needed to induce rotational instability are unlikely.  
The measured brightness variations are probably not due to low-level dust ejection from the nucleus, because this would require ejection speeds of at least 100 m s$^{-1}$ in order for the dust to escape the photometry aperture in such a short time.  This is roughly two orders of magnitude higher than the out-of-plane velocity components derived in Section \ref{sec:dust}.

We next use the faintest absolute magnitude, $H_V$ = 17.0$\pm$0.1 (Table \ref{photometry_keck}), to estimate the parameters of the nucleus.  The relation between the equivalent circular diameter measured in kilometers, $D_{km}$, the visual geometric  albedo \mbox{$p_v$} and the absolute magnitude, $H_V$, is (Harris and Lagerros 2002)

\begin{equation}
D = \frac{1329}{p_v^{1/2}}10^{-H_V/5}.
\label{D}
\end{equation}

\noindent Geometric albedos of most asteroids near $R$ = 3 AU fall in the range 0.04 $\le p_v \le$ 0.1 (c.f. Figure 13 of Masiero et al.~2011).  We adopt $p_V$ = 0.04 and use Equation (\ref{D}) with $H_V$ = 17.0 to find  $D$ = 2.6$\pm$0.1 km as a likely upper limit to the effective diameter. Under the assumption of a bulk density  \mbox{$\rho$ = 2000 kg m$^{-3}$} and a spherical shape, the approximate gravitational escape speed from 288P is $V_e \sim$ 1.4 m s$^{-1}$.
 
We use \mbox{$\rho$ = 2000 kg m$^{-3}$} for both nucleus and dust throughout this paper. This is an average value for asteroids, and may vary up to 50\% in individual objects (Britt et al., 2002). There is no strong reason to assume that nucleus and dust have the same bulk density, but lacking detailed information on the dust properties we prefer to keep the assumptions as simple as possible. The uncertainty of the bulk density translates to a 50\% uncertainty in our derived nucleus and dust masses, and to a $\sim$20\% uncertainty in the nucleus escape speed.

Our derived escape speed is significantly higher than the value of 0.2 m  s$^{-1}$ derived by Licandro et al.~(2013) from dynamical modelling of the dust motion. The difference is that our value represents the gravitational escape speed of a non-rotating body having the size of 288P, while Licandro et al.~(2013) derived the actual minimum speed of the escaping particles. The latter can be lower than the nominal escape speed due to deceleration inside the asteroid's Hill sphere, to fast rotation of the nucleus, to a strongly aspherical shape, or a combination of these.

\subsection{Dust}
The dust in the Hubble images of 288P occupies an extraordinarily thin sheet extending to either side of the nucleus (Figure \ref{image_hst}). The west arm lies nearly along the direction of the projected negative orbital velocity vector, suggesting that it consists of slow-moving, presumably large particles ejected long before the HST observations. It extends beyond the edge of the field view, an angular distance of 40$\arcsec$ from the nucleus and a linear distance of 50,000 km in the plane of the sky. 
The east arm extends roughly to the anti-sun direction, suggesting that it consists of recently released, small particles accelerated away from the nucleus by radiation pressure.  

\begin{figure}
\epsscale{1.0}
\begin{center}
\plotone{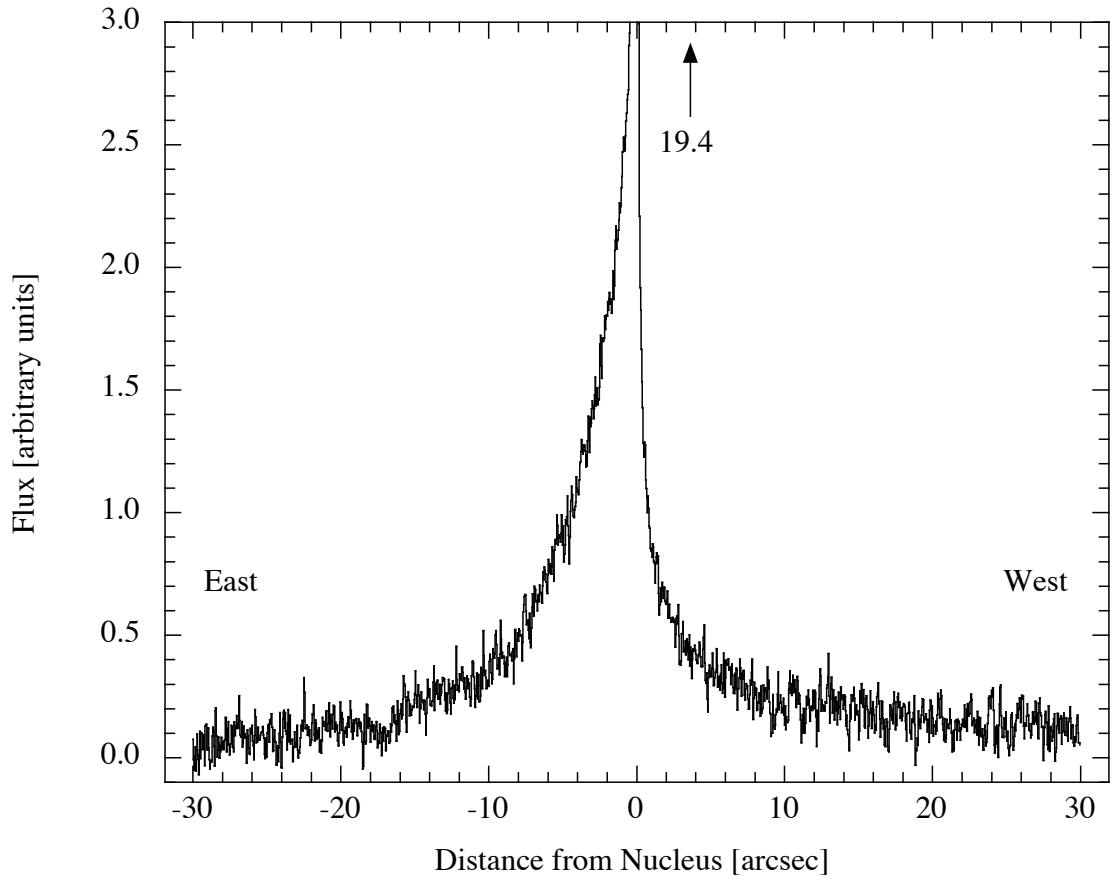}
\caption{Surface brightness profile as a function of position along the dust trail, with the nucleus located at $x$ = 0.  Surface brightness is given in arbitrary units.   \label{profiles} } 
\end{center} 
\end{figure}

We rotated the drizzle-combined average HST images to align the projected orbit of the nucleus with the $x$-axis, and extracted flux profiles perpendicular to the projected orbit. At larger nucleus distances, we averaged over segments of up to 100 pixels (4 arcsec) parallel to the projected orbit in order to obtain a meaningful signal-to-noise ratio. The surface brightness measured along the axis shows that the dust arms are asymmetric, with a steeper drop in surface brightness to the west than to the east (Figure \ref{profiles}).

\begin{figure}
\epsscale{.80}
\plotone{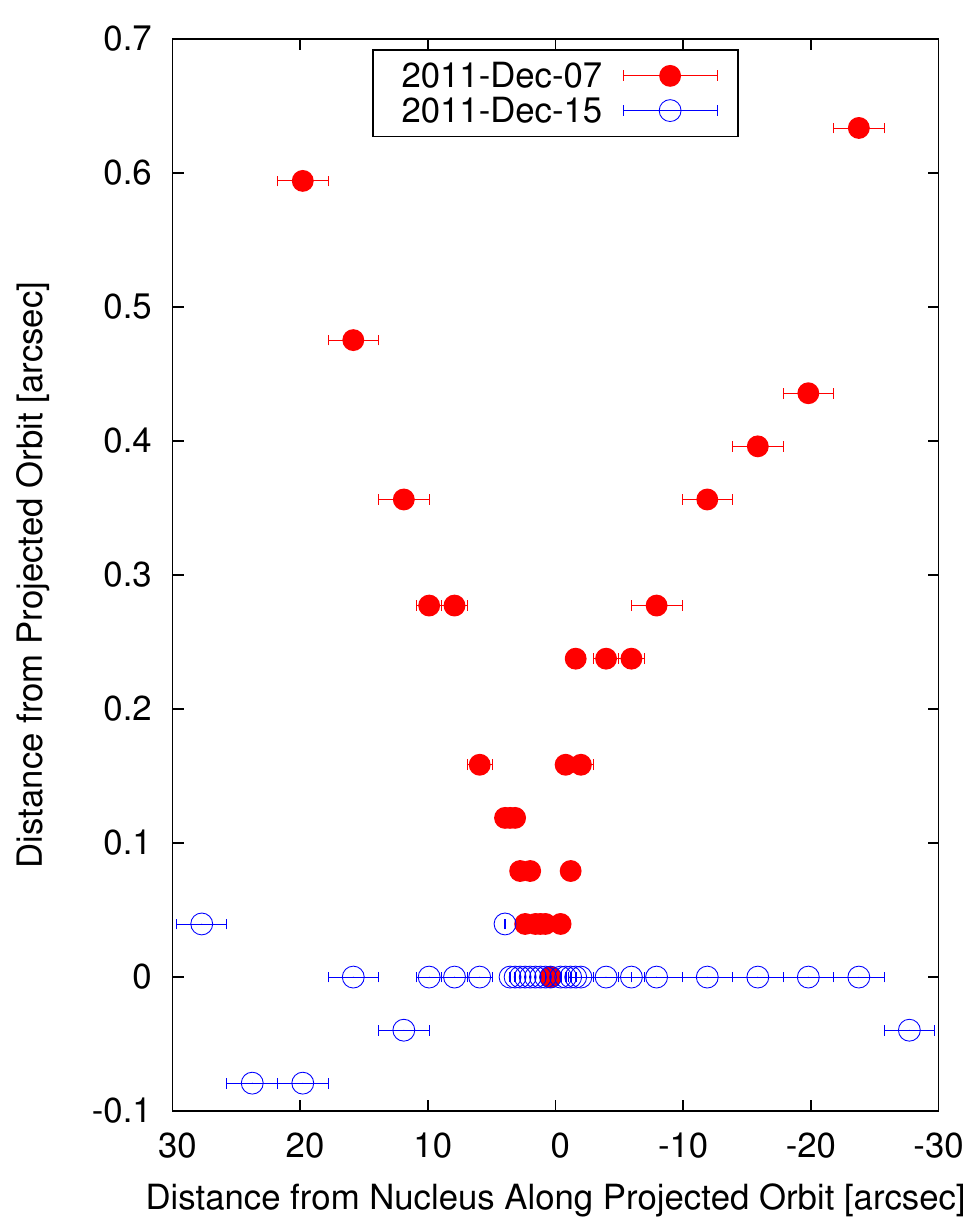}
\caption{Position of the cross-section peak in perpendicular tail profiles as a function of distance from the nucleus on 2011 December 07 and 15 and relative to the projected orbit. Positive values of x indicate positions east of the nucleus, positive values of y refer to positions north of the projected orbit. The error bars indicate the range over which the image was averaged parallel to the projected orbit before measuring the profiles.}
\label{fig:peak}
\end{figure}

Figure~\ref{fig:peak} shows the position of the peak in perpendicular tail cross-sections in the composite images as a function of nucleus distance. On December 15, when the Earth was in the orbital plane of 288P, this peak is close to the projected orbit, showing that the dust was concentrated in the orbital plane of the nucleus. On December 07, when the Earth was displaced from the orbit plane of 288P by 0.27\degr, we see a significant offset between the cross-section peak and the projected orbit that increases with nucleus distance. Combining the in-plane and out-of-plane perspectives, we infer that the brightest axis of the tail is located in the orbital plane but displaced from the orbital path to the direction outside the orbit. 
This implies a spatial separation between the inner edge of the dust sheet and the orbital path within the orbital plane.

\begin{figure}
\epsscale{.80}
\plotone{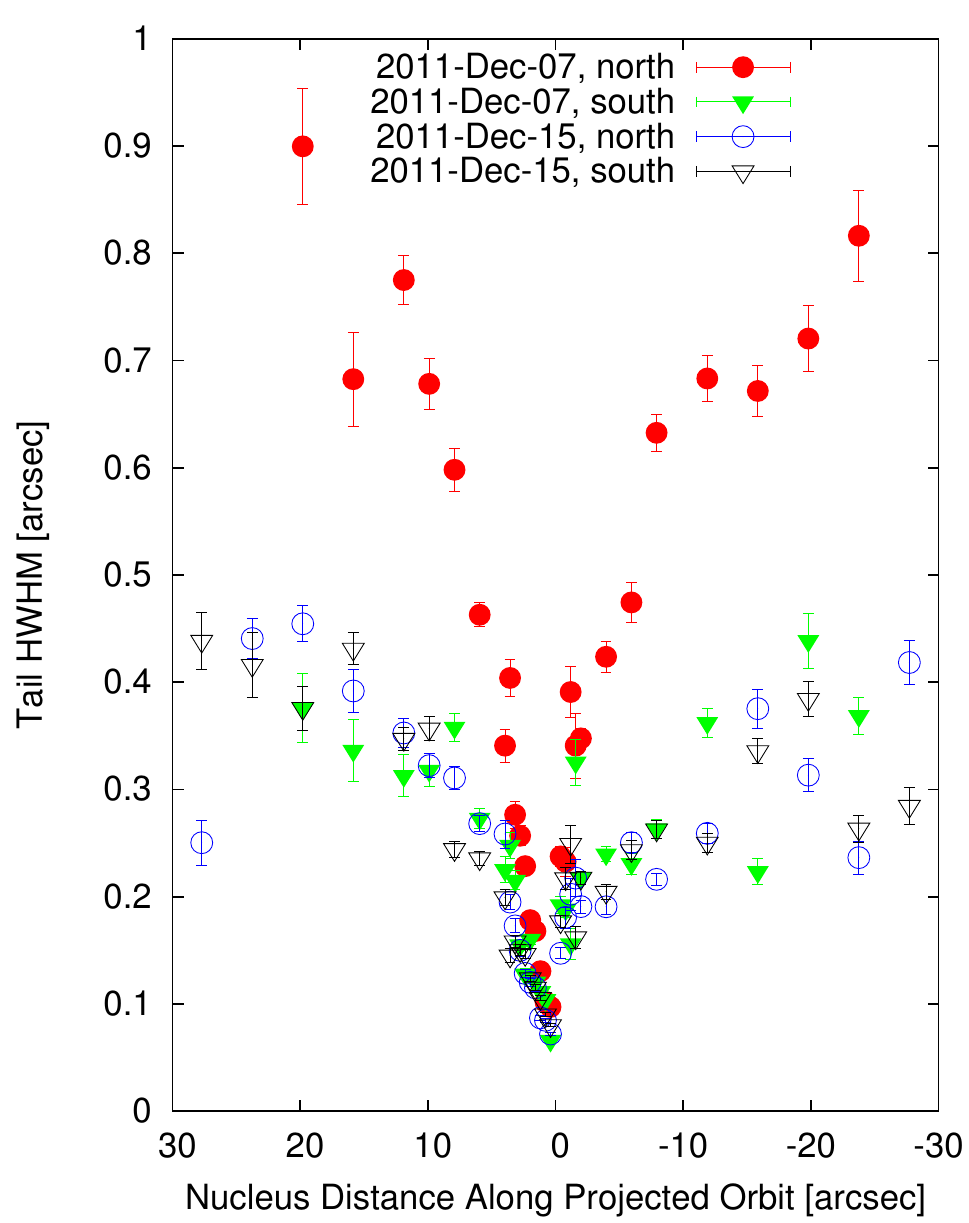}
\caption{Width of the tail (Gaussian standard deviation) north and south of the cross-section peak, and on both HST observation dates. }
\label{fig:width}
\end{figure}

We fitted Gaussian functions to the profiles separately for the northern and southern flanks, keeping the center fixed to the position of the cross-section peak. Figure~\ref{fig:width} shows the HWHM of the tail. 
Both arms become thicker with increasing distance from the nucleus albeit at rates that differ to the east and the west. Subarcsecond HWHM values in Figure (\ref{fig:width}) indicate that the dust in 288P was never spatially resolved in ground-based data (Hsieh et al.~2012, Licandro et al.~2013).
On December 15, the width measures the distribution of dust perpendicular to the orbital plane. There is no significant North-South asymmetry.
On December 07, the southern extent is comparable to that on December 15, suggesting that we see only the out-of-plane extent of the dust. The northern profile, by contrast, is significantly wider on December 07, implying that the dust is spread out in the orbital plane behind the location of the cross-section peak.

In summary, the data are consistent with looking down onto a sharp-edged sheet of dust lying in the orbit plane. The perpendicular extent of the sheet is seen in the December 15 image, while the in-plane distribution of dust can be inferred from the December 07 image. 
We return to the interpretation of the dust morphology in Section~\ref{sec:dust}.

Integrated light photometry of the east and west arms was obtained as follows. 
In the rotated images, we defined two rectangular regions extending $\pm$1.2$\arcsec$ from the mid-plane perpendicular to the dust axes and from 0.2\arcsec~to 30\arcsec~east and west of the nucleus, and measured the total light within each. Background was determined from the average of sky regions located symmetrically above and below the dust tails.
We experimented with larger and smaller boxes and found the above to be ideal in terms of minimizing sky subtraction errors in the photometry. 
In addition, we measured the total light from a 0.44\arcsec~wide strip centered on the nucleus.
 The results are listed in Table (\ref{arms}).  Evidently, while the central region of the comet fades between December 07 and 15, the east and west dust tails do not.    We include in Table (\ref{arms}) estimates of the dust scattering cross-sections computed using the same correction to unit $R$, $\Delta$ and 0\degr~phase angle as for the nucleus (Equation \ref{absmag}) and with $p_V$ = 0.04.  There is no strong reason to expect that the dust albedo and phase function should be the same as those of the nucleus. The volume sampled by our aperture of fixed angular size was 6\% larger on December 15 than on December 07, which does not fully explain the increased cross-section in the dust arms in Table (\ref{arms}), and leads us to underestimate the loss of dust from the central region.

Table (\ref{arms}) shows that the east and west dust arms have scattering cross-sections $C_d(e)$ = 24 km$^2$ and $C_d(w)$ = 13 km$^2$.   The corresponding dust mass, $M_d$, is given by

\begin{equation}
M_d = \frac{4}{3} \rho \overline{a} C_d
\label{mass}
\end{equation}

\noindent where $\rho$ is the dust density (we assume $\rho$ = 2000 kg m$^{-3}$), $\overline{a}$ is the average dust grain radius and $C_d$ is the cross section inferred from the photometry.   We assume a power-law size distribution in which the number of grains having radii between $a$ and $a + da$ is $n(a)da = \gamma a^{-q}da$, and $q$ = 3.5. Such a size distribution describes a collection of particles in collisional equilibrium (Dohnanyi, 1969), and is in agreement with dust size distributions inferred in P/2010 A2 (Jewitt et al., 2010, 2013a, Snodgrass et al., 2010, Kleyna et al., 2013) and 133P (Jewitt et al., 2014b).
With minimum and maximum grain radii $a_-$ and $a_+$, respectively, we compute $\overline{a}$ = $a_+/\ln(a_+/a_.)$ (Jewitt et al., 2014b).  Substituting $a_-$ = 10 $\mu$m and $a_+$ = 300 $\mu$m (see Section~\ref{sec:dust}) we obtain $\overline{a}$ = 88 $\mu$m.  Then, from Equation (\ref{mass}), we estimate the mass of material in the east and west arms as $M(e)$ = 6$\times$10$^6$ kg and $M(w)$ = 3$\times$10$^6$ kg, respectively.  Together, these are $\sim$10$^{-6}$ of the nucleus mass ($M_n$ = 2$\times$10$^{13}$ kg, assuming the same density, albedo and an effective diameter of 2.6 km). Our estimated dust mass is roughly in agreement with results obtained by Licandro et al.~(2013) from ground-based data.

\begin{figure}
\epsscale{0.9}
\begin{center}
\plotone{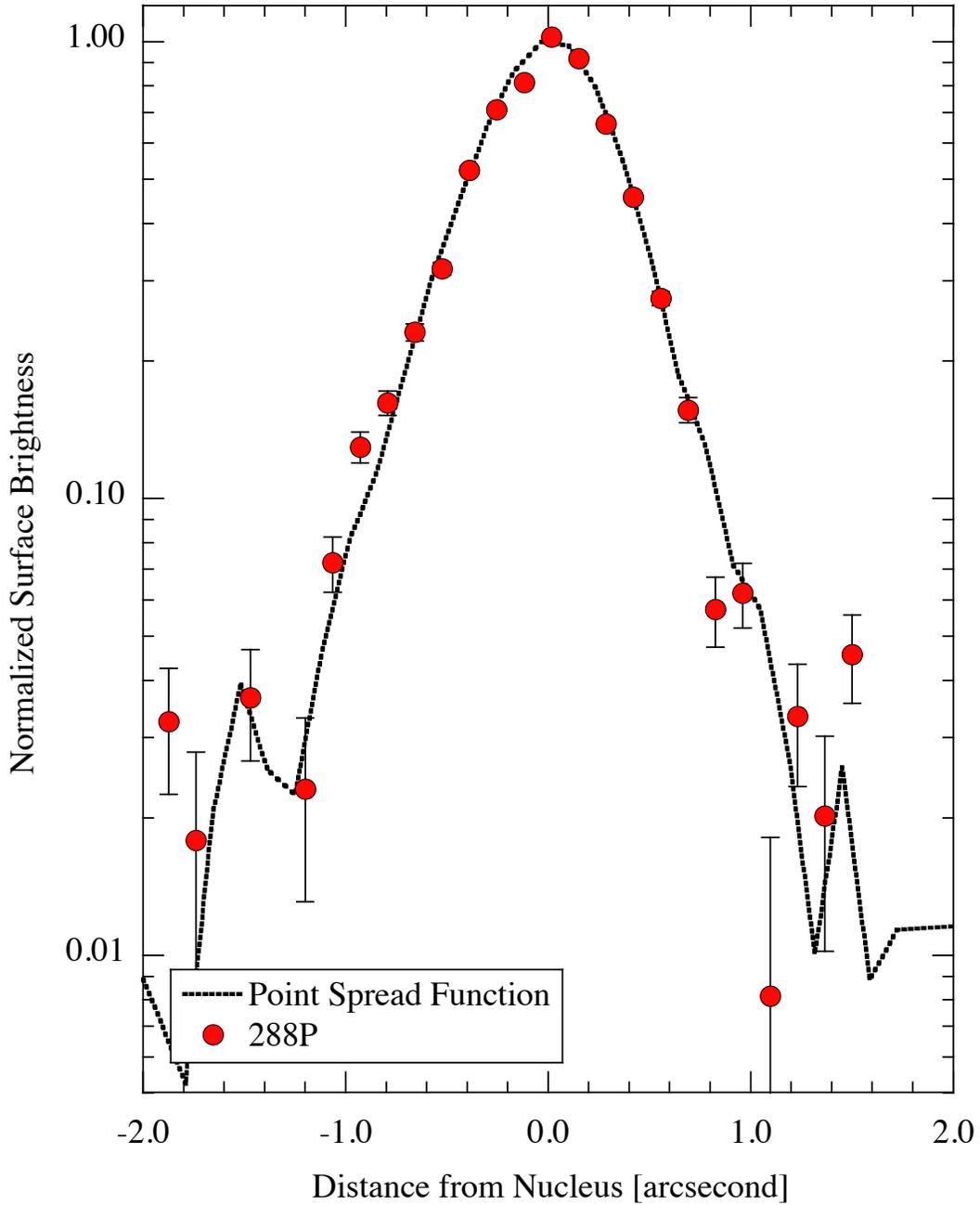}
\caption{Surface brightness profile of 288P (red circles) and the point spread function (black dashed line) in Keck data from UT 2012 October 14.  The profile of 288P is centerd on the object and normalized such that unity corresponds to 22.70 red magnitudes per square arcsec. The stellar PSF was measured in nearby stars from cuts taken perpendicular to the direction of non-sidereal motion projected onto the sky. Error bars correspond to a $\pm$1\% uncertainty in the flat fielding of the data.  \label{SBplot2012Oct14.pdf} } 
\end{center} 
\end{figure}

To conduct a deep search for dust in the UT 2012 October 14 Keck data, we  aligned and combined seven R-band images (each of 300 s exposure) using a median-clipping algorithm.   This had the effect of removing trailed field stars, so providing an uncontaminated image of the asteroid, in which no coma is visually apparent.  In order to conduct a more sensitive search for coma, we compared the surface brightness profile of 288P with the point spread function (PSF) of nearby field stars.  The Keck images were taken using non-sidereal tracking, causing field stars to be trailed by 2.8\arcsec.  For this reason, we measured the surface brightness along cuts taken perpendicular to the trail direction and averaged over 8.1\arcsec~ parallel to the trail direction, in order to capture all of the light.  Object 288P was measured in the same way in order to prevent systematic differences between the profiles.  The results are shown in Figure (\ref{SBplot2012Oct14.pdf}), where the surface brightness has been normalized to 22.70 red magnitudes per square arcsec.  Error bars in Figure (\ref{SBplot2012Oct14.pdf}) show the effect of a $\pm$1\% uncertainty in the flat-fielding of the data.  

We estimated the coma contribution in two ways.  Firstly, we applied the profile convolution model of Luu and Jewitt (1992), assuming a steady-state coma in which the surface brightness falls in inverse proportion to the angular distance from the nucleus of 288P.  We used the PSF from Figure (\ref{SBplot2012Oct14.pdf}) assuming that the function is independent of azimuth.  The model sets a limit to the coma to nucleus brightness ratio of $\sim$10\%, measured within a projected distance of $r$ = 2\arcsec~from the nucleus.  Secondly, we used the relation 

\begin{equation}
m_c = -2.5\log_{10}(2\pi r^2) + \Sigma(r)
\label{JD}
\end{equation}

\noindent from Jewitt and Danielson (1984), appropriate to a steady state coma in which the surface brightness falls inversely with the angular distance from the nucleus.  Here, $m_C$ is the magnitude of a steady state coma integrated out to radius, $r$,  whose surface brightness at $r$ is $\Sigma(r)$ magnitudes~per square arcsec.  A coma with a surface brightness at $r$ = 2\arcsec~systematically more than 1\% of the peak surface brightness would be noticable in our data, corresponding to $\Sigma(2) \ge$ 27.7 magnitudes~per square arcsec.  Substituting in Equation (\ref{JD}), we obtain $m_c >$ 24.2.  For comparison, the measured magnitude of 288P in these data is $m_R$ = 22.45$\pm$0.02, or a brightness ratio of $\sim$5:1.  We conservatively take the less stringent of the convolution and aperture-based estimates to conclude that not more than $\sim$20\% of the light from 288P in the Keck data (corresponding to $\sim$ 1~km$^2$) can be contributed by an unseen, steady-state, near-nucleus coma.  The mass of dust computed from this cross-section by Equation (\ref{mass}) is 2.3 $\times$ 10$^5$ kg.  Of course, this is a model-dependent constraint and comae having surface brightness profiles much steeper than the steady state case could exist undetected, in which case the central dust cross-section could be larger.

\subsection{Near-nucleus region}
\label{subsec:PSF}
\begin{figure}
\epsscale{0.7}
\begin{center}
\plotone{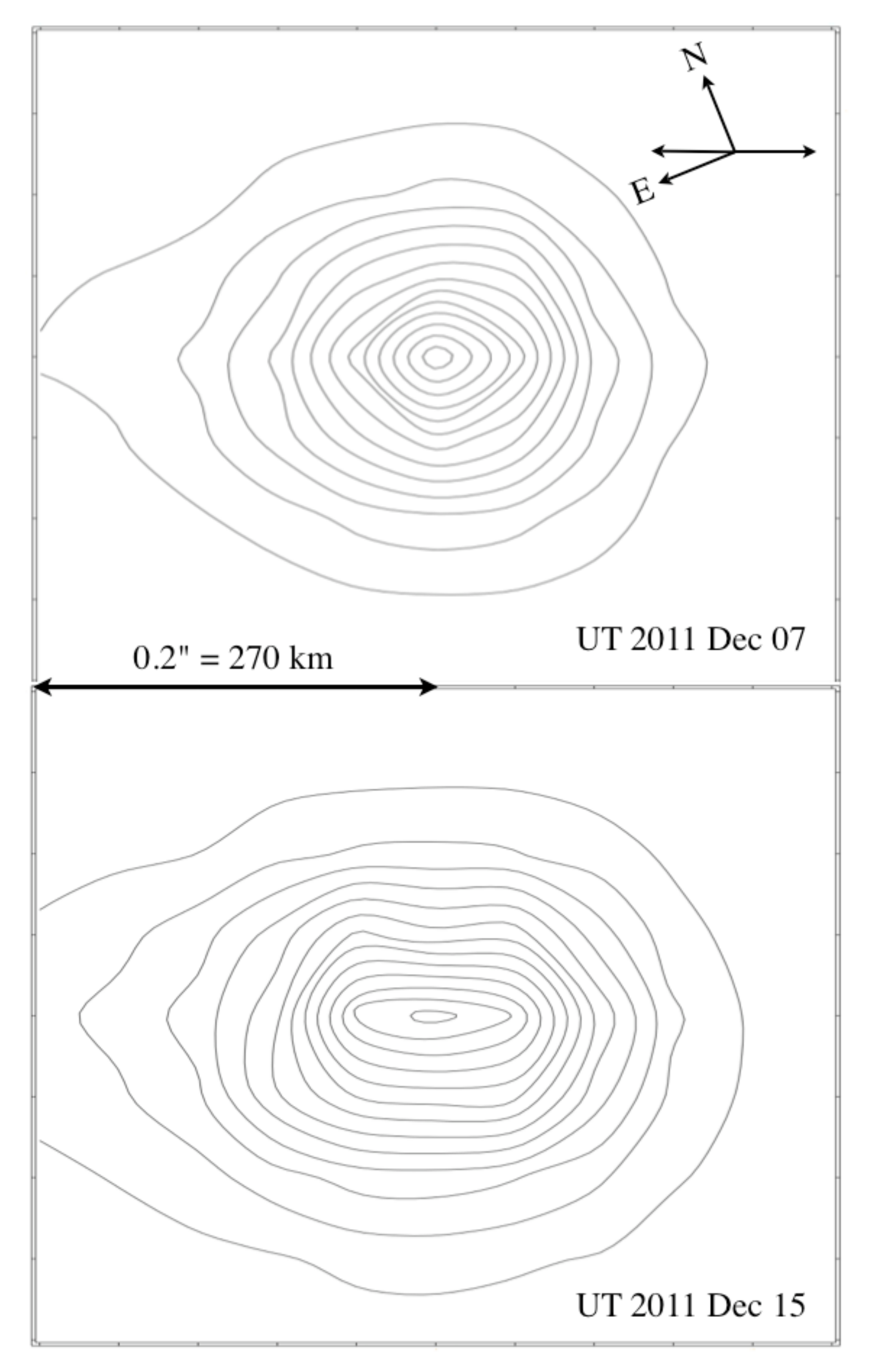}
\caption{Contours of the central regions of the images from UT 2011 December 07 and 15. The innermost contour in each image corresponds to a surface brightness of 16.6 magnitudes per square arcsec, each following contour corresponds to a surface brightness 20\% fainter than the previous one. A 0.2\arcsec~scale bar is shown.     \label{contours} } 
\end{center} 
\end{figure}

Close comparison of the nucleus region reveals a change in the morphology between the two HST observations.  Data from December 07 show a symmetrical nucleus PSF while on December 15 the PSF is strongly elongated parallel to the projected orbit (Figure \ref{contours}).  The FWHM measured along position angle 250\degr~is 0.14$\pm$0.01$\arcsec$ on December 07 rising to 0.19$\pm$0.01$\arcsec$ on December 15.  The difference, while small, is highly significant given the stability of the PSF of HST. The apparent elongation
of the PSF cannot be due to inaccurate alignment of frames in the
composite image, because the elongation is also present in the single
exposures. Neither did we find any indication in the engineering data for a pointing or
tracking error in the December 15 observation. We therefore assume
that the change in isophot shape is not an instrumental artefact. 
Taking into account also the overall fading of the nucleus region (by $\sim$20\% between December 07 and December 15 after geometric correction) and the near-constant brightness of the eastern and western tails, we explore possible explanations. 

One possibility is that the elongation of flux contours on December 15 could be due to additional dust in the orbital plane. The average absolute magnitudes in the central apertures (c.f.~Tables~\ref{photometry_hst} and \ref{photometry_keck}) correspond to cross-sections of $C_{HST1}$ = 8.4 km$^2$, $C_{HST2}$ = 6.9 km$^2$, and $C_{Keck}$ = 5.5 km$^2$. Assuming that the Keck image has only nucleus flux, we estimate that the dust contribution in the central apertures of the HST images is of order 25 to 50\%. 
We performed numerical experiments of combining artificial images of a linear profile (representing the tail) and a point (the nucleus) with various flux ratios and convolving them with a Gaussian function having a standard deviation of 1.681 pixels to simulate the HST/WFC3 PSF at 600nm (Dressel 2015). The linear intensity profile populated only pixels along the central axis and was extrapolated from the tail profiles shown in Figure~\ref{profiles}.
We estimate that, to achieve the observed deformation of the PSF, the ratio of nucleus to tail flux in the central aperture must have decreased from about 2 to 0.5 between December 07 and 15, corresponding to nucleus (dust) contributions of 5.6 (2.8)~km$^2$ on December 07 and 2.3 (4.6)~km$^2$ on December 15. This requires an axis ratio of $>$2.4 if it is assumed that the December 07 (15) observation showed the maximum (minimum) cross sections of a prolate spheroid. The additional dust could be due either to a short-term increase of activity, or to a projection effect caused by the Earth's position in the orbital plane of 288P. The latter would naturally explain why the elongation of the PSF is aligned with the projected orbital plane, but is not supported by the overall fading of the more distant tail between December 07 and 15.

Another possible  explanation for the change in PSF shape is that 288P could have a binary nucleus, barely resolved in the second but not in the first image. The overall fading of flux in the 5 pixel aperture would then be explained by this aperture not covering the full shapes of the PSFs of the two nuclei offset from each other by 1-2 pixels. In addition, dissipation of dust and different rotational phasing remain possible causes of the fading.
Assuming equally sized nuclei with a common cross-section equal to that of a single $D$ = 2.6 km diameter sphere (Section~\ref{subsec:nucleus}), each would have a diameter of $D_b = D/\sqrt2$ = 1.8 km. The Hill sphere of one such body with the orbital elements of 288P has a radius of about $r_H$ = 470 km, about 9 HST pixels. A binary system with mutual distance of 1-2 pixels would therefore be stable.

The semi-major axis and mutual orbital period of the binary system is constrained by the observed projected velocity and the estimated mass of the system. The total mass of two bodies of diameter $D_b$ and a density of 2000\,kg m$^{-3}$ is $M_{2b}$ = 13\,$\times$\,10$^{12}$ kg. With a minimum semi-major axis of 50\,km (1 pixel), the orbital period must be larger than 28 days, hence the HST observations sampled less than half a period. Assuming circular orbits and that the two nuclei were at conjunction during the December 07 observation, the true anomaly $\nu$ on December 15 is given by $\nu = 4 \pi \Delta T_b / T_b(a_b)$, with $\Delta T_b$ = 8 days and $T_b(a_b)$ given by Kepler's third law. The projected distance is $d = a_b \sin \nu$, assuming that the orbital plane of the binary system coincides with the heliocentric orbital plane, which is suggested by the alignment of the elongated PSF with the projected orbit. The relation between $d$ and $a_b$ is given by

\begin{equation}
d(a_b) = a_b \sin \left( 4 \pi \Delta T_b \sqrt{\frac {G M_{2b}}{4 \pi^2 a_b^3}}\right).
\label{eq:binary_period}
\end{equation}
From the shape of the PSF we know that on December 15, 50\,km $< d <$ 150\,km. Plotting Equation~\ref{eq:binary_period} shows that $d <$ 100\,km. 
For $d >$ 80\,km, we find 80\,km $< a_b <$ 250\,km, with a corresponding 
period of 56\,d $< T_b <$ 310\,d. Even a 300\,km separation (0.24\arcsec) is below the seeing-limited resolution of ground-based observations. 

In summary, if 288P is a binary system, we expect the two nuclei to have a separation of order 100 km and to be of similar size, because nuclei of strongly different sizes would not cause the observed deformation of the PSF.
This combination of high ($\sim$1) mass ratio and wide separation is different from known small asteroid binaries, which have either high  mass ratio and small separations (``group B'') or large separations at low mass ratio (``group W''), or neither (e.g. Margot et al., 2015, Walsh et al., 2015). However, it could be due to the difficulty in discovering them that similar systems are not yet known: the long period of the mutual orbit means that eclipsing or occulting events are rare and not easily detected in a lightcurve, while the separation is not large enough for the components to be resolved by a ground-based telescope in a Main Belt object, and observations with highly-resolving telescopes (AO or space-based) are rare, in our case only triggered by the unusual activity of 288P. 
Small binaries typically form by rotational fission of a precursor body (Margot et al., 2015, Walsh et al., 2015). We hypothesize that if 288P is a binary formed by rotational fission, the disruption may have uncovered a spot of primordial ice from the interior of the asteroid that now sublimates when illuminated and causes the observed activity.

Unfortunately, in the absence of additional observations, we possess no way to decide which of the above explanations for the wider PSF on December 15, if either, is correct.

\section{DUST DYNAMICAL ANALYSIS}
\label{sec:dust}
We constrain the size range, velocities, and ejection times of dust ejected from 288P from models of the tail. The trajectory of a dust particle is determined by its initial velocity on leaving the nucleus, and by $\beta$, the ratio of solar radiation pressure acceleration to local solar gravity. For homogeneous spheres of radius, $a$, and  bulk density, $\rho$,  $\beta = 5.77 \times 10^{-4} Q_{\rm pr}/(\rho a)$, where the dimensionless parameter $Q_{\rm pr}$ characterises the optical properties of the material (Burns et al., 1979).
Radiation pressure acts like a mass-spectrometer in a small body's dust tail, allowing the size of a particle to be inferred from its orbital evolution. 

The non-detection of dust within 90\arcsec\ of the nucleus on 2012 October 14 distinguishes 288P from, for example, the active asteroid P/2010 A2, that was still embedded in a trail of cm-sized particles more than 4 years after the disruption event (Jewitt et al., 2013a). The lack of a dust trail suggests that a negligible quantity of such large particles was ejected, and may hint at a different ejection mechanism.
We derive a lower limit for the radiation pressure parameter $\beta$ from the calculated positions of test particles having a wide range of initial velocities, ejection times and $\beta$. 
We consider values of  $0 \leq \beta \leq 10^{-3}$ in steps of $10^{-5}$. The two Cartesian, in-plane velocity components independently range from --1 m s$^{-1}$ to +1 m s$^{-1}$ in steps of 0.01 m s$^{-1}$. The velocity component perpendicular to the orbital plane is kept zero, because it barely influences the motion of grains parallel to the orbital plane. 
We study ejection dates ranging from 2011 March 20 (120 days before perihelion) to 2011 November 15 (120 days after perihelion), in intervals of 30 days. 
We find no test particle located more than 90\arcsec\ east of the nucleus. In the west, the minimum $\beta$ compatible with the 90\arcsec\ limit from the Keck deep image depends on both ejection velocity and date. For zero ejection speed, the lower limit for $\beta$ ranges from 1.5\,$\times$\,10$^{-4}$ for ejection in 2011 March, to 6\,$\times$\,10$^{-4}$ for ejection in 2011 November. Non-zero ejection speeds below 1.4 m s$^{-1}$ (our tested range) offset the minimum value of $\beta$ by up to 2.6\,$\times$\,10$^{-4}$. We conclude that no particles having $\beta <$ 10$^{-4}$ were ejected, which corresponds to a maximum size of 3~mm  for a bulk density of 2000\,kg m$^{-3}$.

To analyse the dust tail in the HST images, we use the concept of synchrones and syndynes (Finson \& Probstein (1968)). A synchrone comprises the loci of particles with variable $\beta$ ejected with zero velocity at a single point in time, while particles on a syndyne are characterised by a constant value of $\beta$ with variable ejection times. 
To evaluate the validity of the zero velocity approximation, we compare the relative influence of initial velocity and radiation pressure on a particle's energy $E$, which by Kepler's law determines its orbital period:

\begin{equation}
\frac{E}{m}
= \frac{1}{2}(\vec{v}_{\rm n} + \vec{v}_{\rm ej})^2 -  \frac{G M_{\odot} (1-\beta)}{r} = \frac{E_{\rm n}}{m_{\rm n}} + \,\vec{v}_{\rm n} \vec{v}_{\rm ej} 
+ \frac{1}{2} \vec{v}_{\rm ej}^2
+ \frac{G M_{\odot} \beta}{r}, 
\label{eq:orb_en}
\end{equation}

\noindent where $r$ is the distance from the Sun, $G$ the gravitational constant, $m$, $m_n$, and $M_{\odot}$ are the masses of particle, nucleus, and Sun, $\vec{v}_{\rm n}$ is the nucleus orbital velocity, $\vec{v}_{\rm ej}$ is the particle's ejection velocity relative to the nucleus, and $E_{\rm n}$ is the orbital energy of the nucleus.
We assume that the zero-velocity approximation is valid if $\vec{v}_{\rm n} \vec{v}_{\rm ej}$ is small compared to the radiation pressure term:

\begin{equation}
v_{\rm ej} << \beta \, \frac{G M_{\odot}}{r \, v_{\rm n}} = \beta \times 1.8 \times 10^4 \, \rm {m\, s}^{-1},
\label{eq:fp_condition}
\end{equation}

\noindent 
with $v_{\rm n} \sim$ 2$\times$10$^4$ m s$^{-1}$, and $r \sim$ 2.5 AU. For $\beta > 10^{-4}$, the zero velocity approximation is valid for particles having velocities well below 1.8 m s$^{-1}$. Further down we show that these conditions are likely fulfilled by the dust seen in the HST images.

\begin{figure}
\epsscale{.80}
\plotone{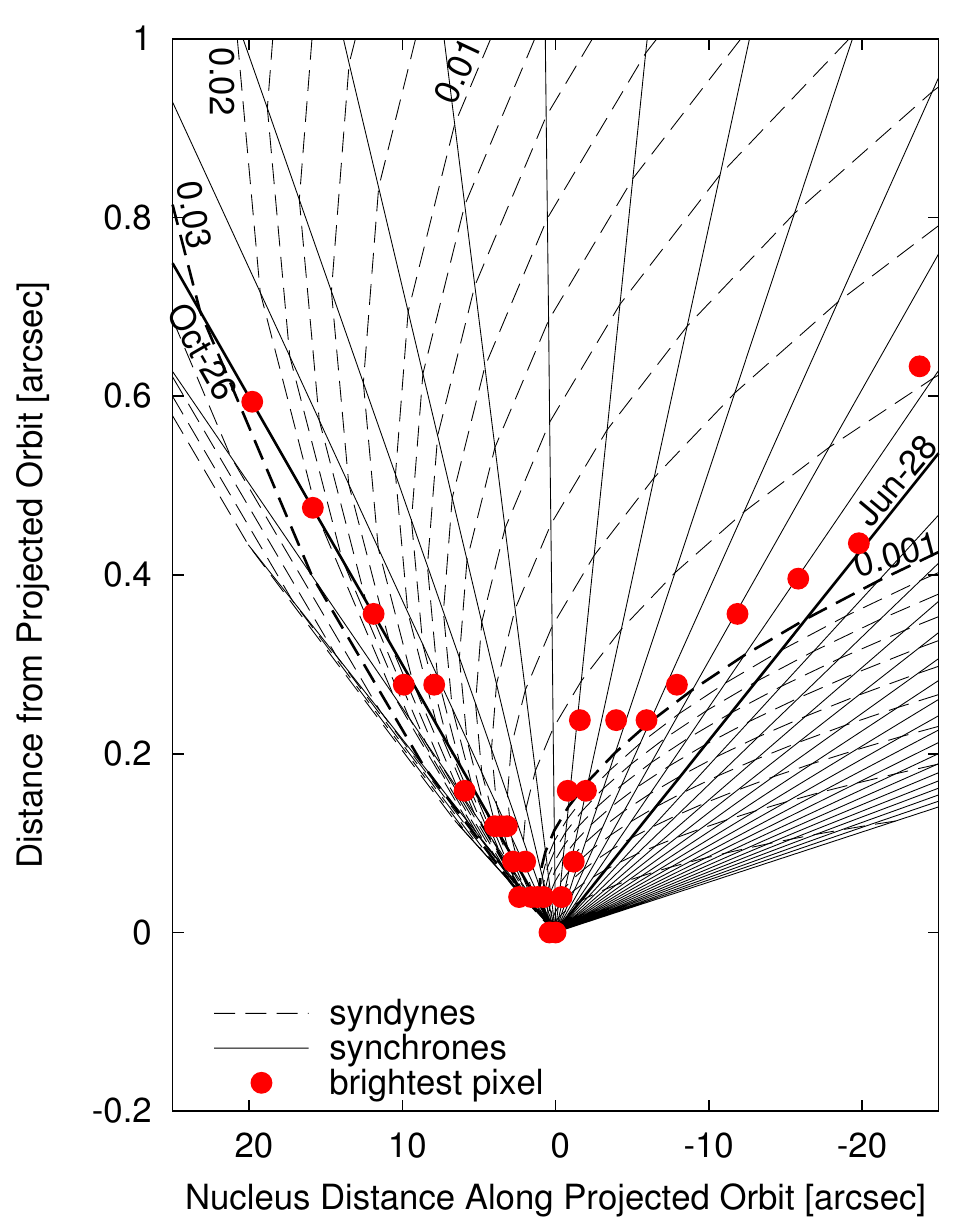}
\caption{Comparison of the location of the cross-section peak on 2011 December 07 to a grid of synchrones and syndynes. Synchrones refer to ejection dates in 2011, and are in steps of 10 days. The numbers marking syndynes refer to the parameter $\beta$.}
\label{fig:synsyn_peak_111207}
\end{figure}

Figure~\ref{fig:synsyn_peak_111207} shows the positions of the cross-section peak (c.f.~Figure~\ref{fig:peak}) together with synchrones and syndynes. East of the nucleus, the peak is located roughly on the synchrone of 2011 October 26, i.e. the bulk of dust seen east of the nucleus was likely ejected around that date. The easternmost extent of the detected tail corresponds to particles having $\beta = 0.025$ (12\,$\mu$m).
West of the nucleus, the peak follows roughly the syndyne of $\beta = 0.001$ (290 $\mu$m). Since the peak defines the inner edge of the dust sheet, we infer that the activity decreased significantly after 2011 October, and that there is no significant amount of dust having $\beta < 0.001$, which is consistent with the lower limit inferred from the 2012 Keck image. 
The western edge of the HST images corresponds to particles ejected in  early 2011 July, which therefore is our upper limit to the onset of activity.

\begin{figure}
\epsscale{.80}
\plotone{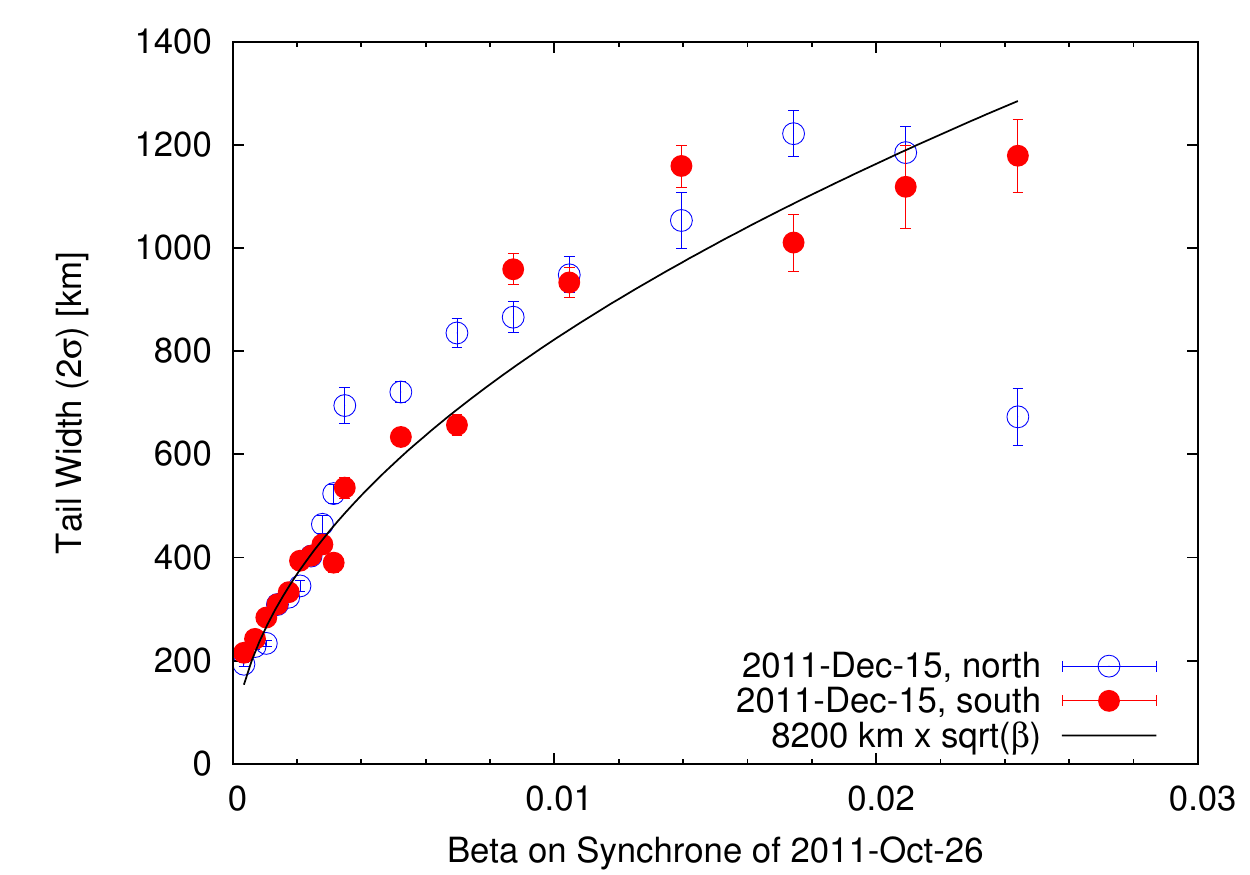}
\caption{Tail one-sided width (2x Gaussian standard deviation) on 2011 December 15 east of the nucleus as a function of the radiation pressure parameter $\beta$. The width was measured as a function of nucleus distance, which was translated to $\beta$ on the assumption that the flux is dominated by particles ejected around 2011 October 26 (see Figure~\ref{fig:synsyn_peak_111207}). 
The solid line is a fit of the relation between width (i.e. velocity) and $\beta$ with a square-root function $v(\beta) = 8198\,{\rm km}/50\, {\rm days} \, \sqrt{\beta} = 1.9\,{\rm m s}^{-1} \, \sqrt{\beta}$. Velocities in the observed range of $\beta$ lie between 0.06 m s$^{-1}$ ($\beta$ = 0.001) and 0.3 m s$^{-1}$ ($\beta$ = 0.025).
}
\label{fig:eastern_tail}
\end{figure}

The width of the tail allows us to constrain the ejection velocities perpendicular to the orbital plane.
Figure~\ref{fig:eastern_tail} shows the eastern tail width (2$\times$ the Gaussian standard deviation) on December 15 as a function of the radiation pressure parameter $\beta$. The width was measured as a function of nucleus distance $x$ (in arcsec), which was translated to $\beta$ according to the linear relation between $x$ and $\beta$ on the synchrone of October 26, $\beta=8.8 \times 10^{-4} x$. The relation between width and $\beta$ is well represented by a square-root function. 
Assuming that the flux in the eastern tail is dominated by dust ejected around October 26 (50 days before the observation), the width is proportional to the perpendicular ejection speed, and we find

\begin{equation}
v_{\perp} (\beta) = (1.9 \pm 0.1)  \sqrt{\beta}.
\label{eq:v_vs_beta}
\end{equation}

\noindent with $v_{\perp}$ measured in ${\rm m~s}^{-1}$. For the largest and slowest particles we find $v_\perp (\beta = 0.001) = 0.06$ m s$^{-1}$. For the smallest particles observed to the east of the nucleus we find $v_\perp (\beta = 0.025) = 0.3$ m s$^{-1}$. Even smaller particles may have been ejected from the asteroid but would have traveled beyond the edge of the FOV and therefore be invisible.

A square-root relation between $\beta$ and ejection velocity is characteristic for dust accelerated by sublimating gas. 
To our knowledge, this is the first observation of an active asteroid from which such a relation could be derived directly from the data. Most often, a square-root relation is assumed a priori by modellers (e.g. Licandro et al., 2013, Jewitt et al., 2014b) based on theoretical studies of comets (Gombosi, 1986). 

\begin{figure}
\epsscale{.80}
\plotone{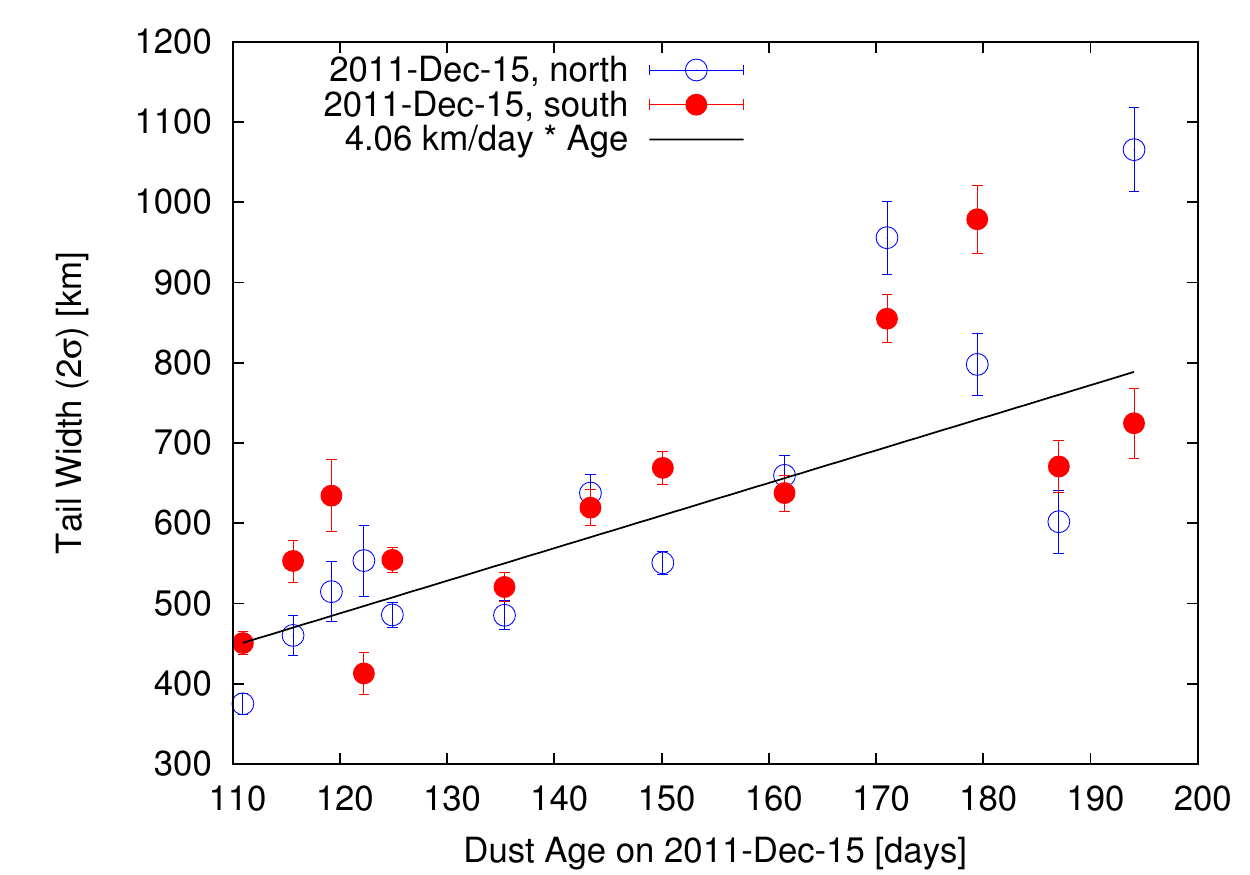}
\caption{
Tail one-sided width (2x Gaussian standard deviation) on 2011 December 15 west of the nucleus as a function of the ejection time. The width was measured as a function of nucleus distance, which was translated to ejection time on the assumption that the flux is dominated by particles having the radiation pressure parameter $\beta$=0.001 (see Figure~\ref{fig:synsyn_peak_111207}). 
We fit the width as a function of age with a linear relationship, the slope of which gives the velocity of the dominant grains ($\beta$ = 0.001). We find that these grains have been ejected at a velocity of 4 km/day, which corresponds to 0.05 m s$^{-1}$.
}
\label{fig:western_tail}
\end{figure}

Figure~\ref{fig:western_tail} shows the width of the western tail on December 15 as a function of ejection time. 
The conversion from distance to ejection time used the relation between these quantities on the syndyne of $\beta$=0.001.
Despite the considerable scatter of the data, we fit a linear function to the width as function of age:
\begin{equation}
w = v_\perp \times (T_{\rm obs} - T_{\rm ej})
\label{fig:width_vs_time}
\end{equation} 
The slope of this function corresponds to the perpendicular ejection velocity of the dominant particles ($\beta = 0.001$), $v_\perp$ = (5.4 $\pm$ 0.2) cm s$^{-1}$. This is roughly in agreement with the speeds inferred from the eastern tail described by Equation~\ref{eq:v_vs_beta}, and implies that the ejection velocity did not significantly change with time.
While the perpendicular velocity is well determined, we cannot constrain the dust velocity within the orbital plane. Assuming that its magnitude is comparable to the perpendicular speed, our inferred ejection times and minimum $\beta$ justify the use of a synchrone-syndyne model.

Our results confirm the inverse square-root relationship between particle size and ejection velocity assumed by Licandro et al.~(2013), and are consistent with their derived maximum grain size of 200 $\mu$m. However, our derived minimum ejection speed (0.06 m s$^{-1}$) is much lower than their 0.2 m s$^{-1}$. One possible reason is that we derive only the out-of-plane component of the velocity. Another possible reason lies in the comparatively low spatial resolution of the ground-based images used by Licandro et al.~(2013) (310 km/pixel compared to 52 km/pixel in the HST images), which may have led them to overestimate the width of the dust sheet and hence the grain velocity.

\section{DISCUSSION}
We seek to use the morphology of the dust in 288P to understand the mechanism behind its ejection.  The key observational results are 

\begin{itemize}
\item The nucleus is a C-type object typical of the outer asteroid belt, with a diameter $D \sim$ 2.6 km (assumed geometric albedo of $p_V$ = 0.04), and corresponding escape speed of 1.4 m s$^{-1}$. 
\item The nucleus brightness varies but shows no evidence for (two-peaked) periods $<$2 hr that would be expected if the nucleus were shedding mass rotationally.
\item Sub-arcsecond changes in the central isophotes between UT 2011 December 07 and December 15 may be evidence for the continued ejection of dust, or indicate that 288P has a binary nucleus with separation of order 100 km.
\item Activity started no later than early in 2011 July (at true anomaly angle of $\nu$ = 355$^{\circ}$) and likely decreased after 2011 October ($\nu$ = 30$^{\circ}$). In October 2012 (at $\nu$ = 107$^{\circ}$), no sign of activity was detected. 
\item The observed dust particles had a radiation pressure parameter in the range 0.001 $< \beta <$ 0.025 (corresponding to sizes between 10 and 300 $\mu$m for a bulk density of 2000 kg/m$^3$). Smaller particles may have been ejected, but our observations were not sensitive to them.
\item The maximum ejection speeds perpendicular to the orbital plane were characterised by the relation $v_\perp$ = 1.9 $\sqrt{\beta}$, and ranged from 0.06 m s$^{-1}$ to 0.3 m s$^{-1}$, well below the nucleus escape speed. Between 2011 July and October, the ejection velocity seems not to have changed with time. 
\item The dust mass in 2011 December is $\sim 10^7$\,kg, only $\sim$10$^{-6}$ of the nucleus mass.

\end{itemize}

Mechanisms suspected to trigger activity in asteroids include rotational disruption, collision with a second asteroid, thermal disintegration, sublimation of ice, and electrostatic forces (Jewitt 2012, Jewitt et al., 2015c). 
Electrostatic forces typically lift particles of a narrow size range that depends on the size of the nucleus. For a 2.6 km-nucleus, the expected particle size is of order 1.5\,$\mu$m, much smaller than observed. Electrostatic processes therefore can be excluded as a cause of the activity in 288P.
Likewise, thermal disintegration is expected to lift primarily very small particles rather than the observed 10 -- 300 $\mu$m-sized ones. 
An impact would cause activity of very short duration, that would manifest itself in a dust tail following a single synchrone. In contrast, the observed dust tail stems from activity continuing over several months and is therefore not consistent with an impact origin.
Rotational disruption occurs as the centrifugal force compensates or exceeds gravity at a body's equator, lifting material from the surface. This process is independent of particle size, and therefore not consistent with the observed absence of particles larger than 300\,$\mu$m. In addition, the velocity in this process should be independent of particle size, while we find an inverse square-root relation. We conclude that super-critical rotation is not the primary trigger of activity in 288P.

Ice sublimation typically causes mass loss sustained over several weeks to months, likely due to seasonal exposure of an icy patch on the surface to sunlight. Dust accelerated by gas drag shows a characteristic inverse square-root relationship between particle size and velocity (e.g. Gombosi, 1986). Both these characteristics are evident in 288P, leading us to think that sublimation of ice is the most likely cause of its mass loss, consistent with earlier findings by Hsieh et al.~(2012) and Licandro et al.~(2013). 

We find that 288P was active in the true anomaly range from at least 355$^{\circ}$ to 30$^{\circ}$, similar to the active orbit sections of the four asteroids with confirmed repetitive activity, which all cover up to a few tens of degrees before and after perihelion. At the orbital position where 288P appeared inactive (at 107$^{\circ}$ true anomaly), 313P has been reported inactive as well (Hsieh et al., 2015), while 324P was still active (Hsieh and Sheppard, 2015). 133P and 238P have not been observed at this orbital position.

However, the observed particle speeds are below the gravitational escape speed of a non-rotating body having the size of 288P. The magnitude of the velocities is comparable to those found in the active asteroid 133P/Elst-Pizarro (Jewitt et al., 2014b), but two orders of magnitude smaller than found in comets of similar size (e.g. Agarwal et al., 2007). Like in 133P, the low ejection velocities can be due to weak activity in a small source on the surface of the asteroid, where horizontal expansion of the gas significantly reduces its final vertical speed as compared to a comet that is active on a more global scale. Also like in 133P, the escape of dust from the nucleus gravity field may be aided by fast rotation reducing the effective escape speed. This is not excluded by our non-detection of a clear 2h-periodicity in the lightcurve, as a fast rotating body may have a flat lightcurve due to unfavourable viewing geometry or near-spheroidal shape.  While sublimation offers the most plausible explanation for the activity in 288P, we regard it as a less convincing case than the four main-belt objects (133P, 313P, 238P, and 324P) in which activity has been observed to reoccur in different orbits. 288P will next reach perihelion on 2016 November 08, and observations around this time should be taken  to search for the recurrence of activity in 288P that is expected if our present interpretation of its origin is correct.

\acknowledgements
Based in part on observations made with the NASA/ESA \emph{Hubble Space Telescope,}
with data obtained from the archive at the Space Telescope Science Institute (STScI). 
STScI is operated by the association of Universities for Research in Astronomy, Inc. 
under NASA contract NAS~5-26555.   Some of the data presented herein were obtained at the W.M. Keck Observatory, which is operated as a scientific partnership among the California Institute of Technology, the University of California and the National Aeronautics and Space Administration. The Observatory was made possible by the generous financial support of the W.M. Keck Foundation.   This work was supported, in part, by a NASA Solar System Observations grant to DJ.

\clearpage

\begin{deluxetable}{lllllllcr}
\tablecaption{Journal of Observations
\label{geometry}}
\tablewidth{0pt}
\tablehead{
\colhead{Instrument} &\colhead{UT\tablenotemark{a}}   & \colhead{$R$\tablenotemark{b}} & \colhead{$\Delta$\tablenotemark{c}}   & \colhead{$\alpha$\tablenotemark{d}} & \colhead{$\theta_i$\tablenotemark{e}} & \colhead{$PA_{Sun}$\tablenotemark{f}} & \colhead{$PA_{V}$\tablenotemark{g}} & \colhead{$\nu$\tablenotemark{h}}}
\startdata
WFC3/F606W & 2011 Dec 07.273 &              2.533  &    1.755   & 16.5 & 0.27 & 66.44 & 247.41 & 39.2\\
WFC3/F606W & 2011 Dec 15.857 &                    2.545   &    1.853   & 18.6 & -0.01 & 67.38 & 247.36 & 41.3 \\
Keck 10 m       & 2012 Oct 14.6 &               3.111 & 3.273 & 17.7 & 0.87 & 286.07 & 0.87 & 107.1\\

\enddata


\tablenotetext{a}{UT mid-date of the observation}
\tablenotetext{b}{Heliocentric distance in AU}
\tablenotetext{c}{Geocentric distance in AU}
\tablenotetext{d}{Phase (sun-object-Earth) angle in degrees}
\tablenotetext{e}{Out-of-plane angle in degrees}
\tablenotetext{f}{Position angle of the anti-solar direction in degrees}
\tablenotetext{g}{Position angle of the projected negative orbital velocity vector in degrees}
\tablenotetext{h}{True anomaly angle in degrees}
\end{deluxetable}

\clearpage

\begin{deluxetable}{llc}
\tablecaption{HST Nucleus Photometry
\label{photometry_hst}}
\tablewidth{0pt}
\tablehead{
\colhead{UT Date 2011} &\colhead{$m_V$\tablenotemark{a}}   & \colhead{$H_V$\tablenotemark{b}}    }
\startdata
Dec 07.2528 & 20.684 $\pm$0.010  &  16.56 \\
Dec 07.2661& 20.677  $\pm$0.010  &  16.55  \\
Dec 07.2716 & 20.654  $\pm$0.010  &  16.53  \\
Dec 07.2776 & 20.656  $\pm$0.010  &  16.53  \\
Dec 07.2831 & 20.653  $\pm$0.010  &  16.53  \\
Dec 07.2886 & 20.680  $\pm$0.010  &  16.56  \\
Dec 15.8431 & 21.136  $\pm$0.010  &  16.81  \\
Dec 15.8486 & 21.090  $\pm$0.010  &  16.77  \\
Dec 15.8541 & 21.057  $\pm$0.010  &  16.73  \\
Dec 15.8601 & 21.057  $\pm$0.010  &  16.73  \\
Dec 15.8656 & 21.059  $\pm$0.010  &  16.74  \\
Dec 15.8712 & 21.069   $\pm$0.010  &  16.75  \\

\enddata


\tablenotetext{a}{Nucleus $V$-band magnitude measured within a 0.2$\arcsec$ radius aperture.  The apparent V-band magnitude was computed from the observed count rate ``C" (in electrons s$^{-1}$) using $m_V = -2.5 \log C + Z$, where $Z$ =  25.99 for the F606W filter (Kalirai et al.~2009).}
\tablenotetext{b}{Magnitude of the nucleus corrected to unit heliocentric and geocentric distances and 0\degr~phase angle (Equation \ref{absmag}).}

\end{deluxetable}

\clearpage

\begin{deluxetable}{llc}
\tablecaption{Keck Nucleus Photometry
\label{photometry_keck}}
\tablewidth{0pt}
\tablehead{
\colhead{UT Date 2012} &\colhead{$m_R$\tablenotemark{a}}   & \colhead{$H_V$\tablenotemark{b}}    }
\startdata
Oct 14.5903	& 22.61$\pm$0.03 & 17.15 \\
Oct 14.6042	& 22.49$\pm$0.03 & 17.03 \\
Oct 14.6088	& 22.48$\pm$0.03 & 17.02 \\
Oct 14.6135	& 22.47$\pm$0.03 &17.01 \\
Oct 14.6181	& 22.45$\pm$0.03 &16.99 \\
Oct 14.6227	& 22.48$\pm$0.03 &17.02 \\
Oct 14.6274	& 22.42$\pm$0.03 &16.96 \\
Oct 14.6319	& 22.45$\pm$0.03 &16.99 \\

\enddata


\tablenotetext{a}{Nucleus $R$-band magnitude measured from Keck data as described in the text.  }
\tablenotetext{b}{Magnitude of the nucleus computed from $m_R$ assuming $m_V - m_R$ = 0.50 and corrected to unit heliocentric and geocentric distances and 0\degr~phase angle (Equation \ref{absmag}).}

\end{deluxetable}

\clearpage

\begin{deluxetable}{llcc}
\tablecaption{Averaged Dust Photometry
\label{arms}}
\tablewidth{0pt}
\tablehead{
\colhead{Date} &\colhead{East Dust Arm\tablenotemark{a}}   & \colhead{Center Region\tablenotemark{b}}  & \colhead{West Dust Arm\tablenotemark{c}}  }
\startdata
2011 Dec 07 & 19.60$\pm$0.10 (22$\pm$2 km$^2$)\tablenotemark{d}  &  20.48  (9.9 km$^2$)\tablenotemark{d} & 20.25$\pm$0.10 (12$\pm$1 km$^2$)\tablenotemark{d}\\
2011 Dec 15 & 19.64$\pm$0.10 (26$\pm$2 km$^2$)\tablenotemark{d}  &  20.93 (7.9 km$^2$)\tablenotemark{d} & 20.29$\pm$0.10 (14$\pm$1 km$^2$)\tablenotemark{d} \\

\enddata


\tablenotetext{a}{Magnitude of the east arm, within 0.2 to 30$\arcsec$ from the nucleus.}
\tablenotetext{b}{Magnitude measured within a 0.44$\arcsec$ wide strip centered on the nucleus.}
\tablenotetext{c}{Magnitude of the west arm, within 0.2 to 30$\arcsec$ from the nucleus.}
\tablenotetext{d}{Conversion from magnitude to cross-section was done assuming a geometric albedo of 0.04 and the same phase function as used for the nucleus.}

\end{deluxetable}

\clearpage

\clearpage

\clearpage

\clearpage

\clearpage

\clearpage

%
%
\clearpage

\clearpage

\clearpage

\clearpage

\clearpage

\end{document}